\documentclass[12pt,preprint]{emulateapj}
\begin{document}
\shorttitle{Acceleration by Reconnection}
\shortauthors{Lazarian \& Opher}
\title{A model of Acceleration of Anomalous Cosmic Rays by Reconnection in the Heliosheath}

\author{A. Lazarian}
\affil{Dept. of Astronomy, University of Wisconsin,
   Madison, WI53706; lazarian@astro.wisc.edu}
\author{M. Opher}
\affil{Dept. of Physics and Astronomy, George Mason University, 4400 University Drive, Fairfax, VA 22030; mopher@gmu.edu}

\begin{abstract}
We discuss a model of cosmic ray acceleration that accounts for the observations of anomalous cosmic rays by Voyager 1 and 2. The model appeals to fast magnetic reconnection rather than shocks as the driver of acceleration. The ultimate source of energy is associated with magnetic field reversals that occur in the heliosheath. It is expected that the magnetic field reversals will occur throughout the heliosheath, but especially near the heliopause where the flows slows down and diverge in respect to the interstellar wind and also in the boundary sector-in the heliospheric current sheet. While the First Order Fermi acceleration theory within reconnection layers is in its infancy, the predictions do not contradict the available data on anomalous cosmic ray spectra measured by the spacecrafts. We argue that the Voyager data is one of the first pieces of evidence favoring the acceleration within regions of fast magnetic reconnection, which we believe to be a widely spread astrophysical process.
\end{abstract}

\section{Introduction}
Since the crossing of the termination shock (TS) by Voyager 1 (V1) in late 2004 and by Voyager 2 (V2) in mid 2007 it became clear that several paradigms needed to be revised. Among them was the acceleration of particles. Prior to the encounter of the termination shock by V1 the prevailing view was that anomalous cosmic rays (ACRs) were accelerated at the TS by diffusive shock acceleration (DSA) to energies 1-300 MeV/nuc (e.g., Jokipii \& Giacalone, 1998; Cummings \& Stone, 1998). However, with the crossing of the TS by V1 the energy spectrum of ACR did not unroll to the expected source shape: a power-law at lower energies with a roll off at higher energies. After 2004, both the V1 spectrum in the heliosheath and the V2 spectrum upstream the TS, continued to evolve toward the expected source shape. 

To explain this paradox several models were proposed. Among them, McComas \& Schwadron (2006) suggested that at a blunt shock the acceleration site for higher energy ACRs would be at the flanks of the TS, where the injection efficiency would be higher for DSA and connection times of the magnetic field lines to the shock would be longer, allowing acceleration to higher energies. Fisk et al. (2006) on the other hand suggested that stochastic acceleration in the turbulent heliosheath would continue to accelerate ACRs and that the high-energy source region would thus be beyond the TS. Other works, such as Jokipii (2006) and Florinki and Zank (2006) try to explain the deficit of ACRs based on a dynamic termination shock. Jokipii (2006) pointed out that a shock in motion on time scales of the acceleration time of the ACRs, days to months, would cause the spectrum to differ from the expected DSA shape. Florinki \& Zank (2006) calculated the effect of Magnetic Interacting Regions (MIRs) with the Termination Shock on the ACR spectral shape. They show that there is a prolonged period of depressed intensity in mid-energies from a single MIR. Other recent works have recently included stochastic acceleration, as well as other effects (Moraal et al., 2006, 2007; Zhang, 2006; Langner and Potgieter, 2006; Ferreira et al., 2007). It became clear after the crossing of the TS by V2 these models would require adjustments. The observations by V2 indicate for example that a transient did not cause the modulation shape of the V2 spectrum at the time of its TS crossing. When both spacecraft were in the heliosheath in late 2007, the radial gradient in the 13-19 MeV/nuc ions does not appear to be caused by a transient. The 60-74 MeV/nuc ions have no gradient, so no north-south or longitudinal asymmetry is observed in the ACR intensities at the higher energies.

Here we propose an alternative model, which explains the source of ACRs as being in the heliosheath. In what follows, we appeal to magnetic reconnection as a process that can accelerate particles. We explain the origin of the magnetic field reversals that induce magnetic reconnection in heliosheath and heliopause in \S2.

It is known that magnetic fields are well frozen-in into space plasma, which causes magnetic fields and plasma move together. However, as soon as two magnetic flux tubes attempt to cross each other, magnetic field lines of different direction come sufficiently close to each other that the magnetic field rearrangement takes place. This topological rearrangement converts free energy of magnetic field into the energy of plasma motion and plasma heating. Many phenomena, including solar flares, the Earth magnetospheric events, $\gamma$-ray bursts are accepted to be powered by magnetic reconnection, although the detailed mechanism of the process stayed
illusive for a long time (see Biskamp 2000, Priest \& Forbes 2000, Bhattachargee 2004, Zweibel \& Yamada 2009). Recent progress achieved in understanding the nature of reconnection processes allows us to appeal to the process with more confidence. In \S 3 we discuss the magnetic reconnection in the heliosheath and propose a model of reconnection there, which is dominated by turbulence on the  large scales, but exhibits properties of collisionless reconnection on microscales. 

The acceleration of energetic particles by reconnection is a subject in the state of development. In \S4 we provide an outlook at the processes of energetic particle acceleration in reconnection regions and identify the first order Fermi acceleration arising from energetic particle bouncing between the reconnecting fluxes as the dominant acceleration process.

We accept the exploratory nature of this work and compare our proposed solution with the alternative solutions of the ACRs problem in \S5. There we also discuss the existing limitations of our understanding of the complex processes that we invoke in our model and how these uncertainties affect our conclusions.  
  
\begin{figure*}[!t]
\includegraphics[width=0.65\textwidth]{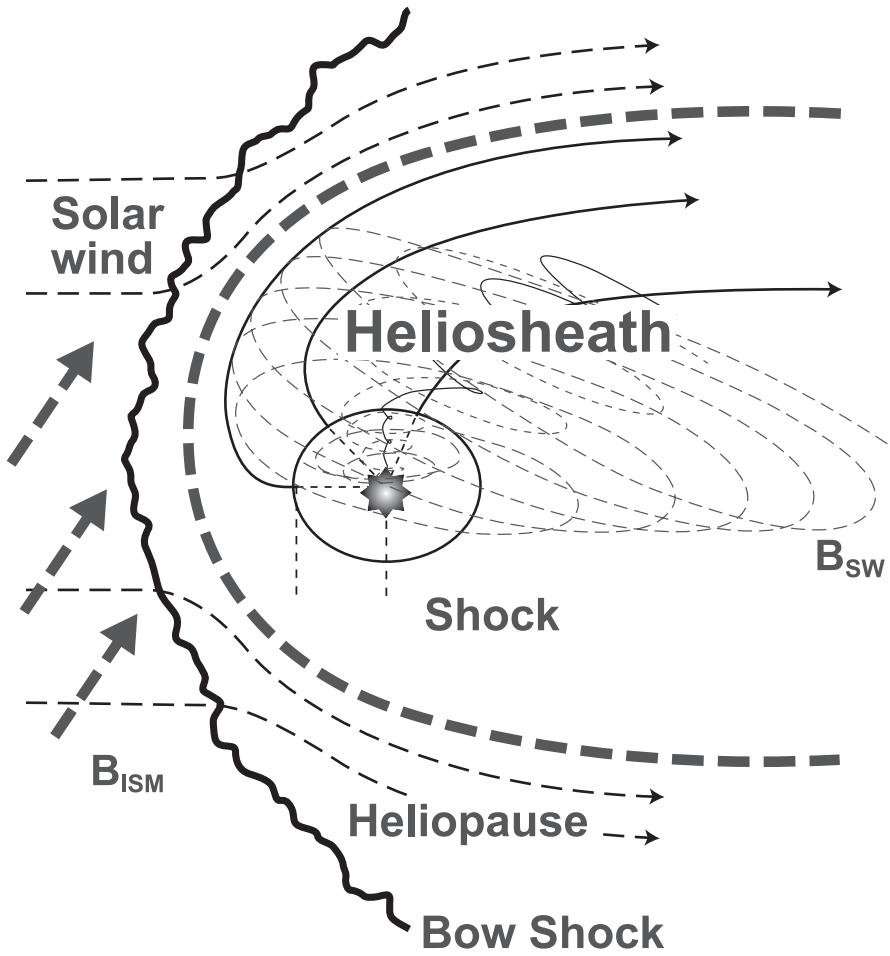}
\caption{Global view of the interaction of the solar wind with the interstellar wind. The spiral solar magnetic field (shown in dark dashed lines) is  shown being deflected at the heliopause. The heliopause itself is being deflected by the interstellar magnetic field. (figure adapted from S. Suess (2006).}
\label{view}
\end{figure*}

\section{Nature of magnetic field reversals expected in the Heliosheath}

It is well known that magnetic field in the heliosphere change 
polarity and create current sheets. For instance,
as the Sun rotates magnetic field twists into a Parker spiral (Parker 
1958) with magnetic fields separated by a current sheet (see Schatten 
1971). The changes of magnetic field are also expected due to the 
Solar cycle activity.

The question now is at what part of the heliosheath we expect to see 
reversals. The structure of the magnetic field in the solar wind is 
complex. The solar magnetic field lines near the termination shock are 
azimuthal and form a spiral (see Figure \ref{view} that shows a 
global view of the interaction of the solar wind with the 
interstellar wind). The spiral solar magnetic field is  shown being 
deflected at the heliopause
(shown in dark dashed lines). The heliopause itself is deflected by 
an interstellar magnetic field being asymmetric noth-south (Opher, 
Stone and Gombosi 2007). It is not clear depending on the intensity 
of the interstellar magnetic field if there is a Bow Shock or not 
(shown as a wiggle line in Figure \ref{view}). The opposite 
hemispheres spirals are separated by the heliospheric current sheet 
(HCS) that is tilted with respect to the solar rotation axis. This 
tilt creates the so called ``baillarina skirt". 

\begin{figure}[!t]
\includegraphics[width=\columnwidth]{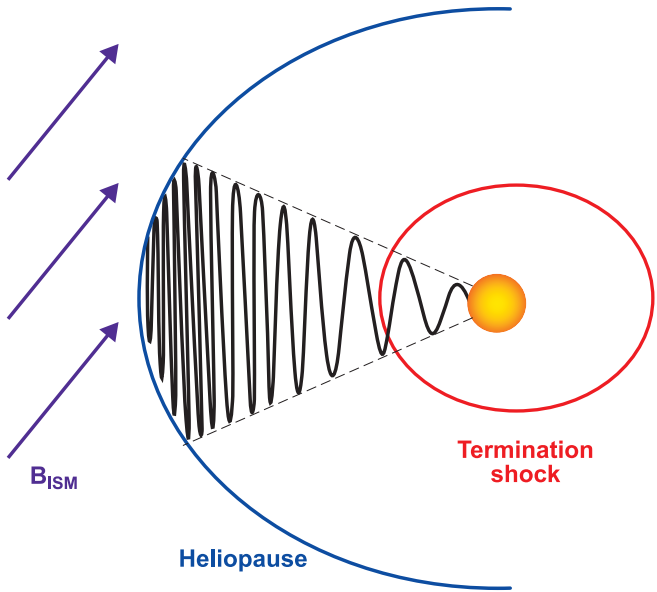}
\caption{ A meridional view of the boundary sectors of the 
heliospheric currenty sheet  and how the opposite sectors get tighter 
closer to the heliopause. The thickness of the outflow regions in the 
reconnection region depends on the level of turbulence. The length of 
the outflow regions $L$ depends on the mean geometry of magnetic 
field and turbulence.}
\label{structure1}
\end{figure}

As the solar 
cycle progresses, the tilt of the HCS increases from close to $10^
\circ$ to be highly inclined. Figure \ref{structure1} shows a side 
(meridional) view of the HCS (black line) for a specific phase of the 
solar cycle, when the tilt is $30^\circ$. The HCS separates opposite 
solar polarities. One can see that before the termination shock, when 
the solar wind is uniform $\sim 400 km/s$, the opposite polarities 
are uniformly separated from each other.

\begin{figure}[!t]
\includegraphics[width=\columnwidth]{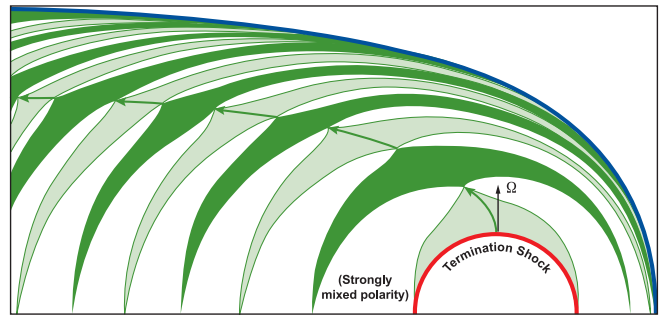}
\caption{A meridional view of the heliosheath showing the effect of 
different solar cycles, represented by consecutive dark and light 
green, as opposite polarities. The rotation axis of the Sun is 
indicated by $\Omega$. The white regions between are strongly mixed 
polarity regions where the effect of variable solar magnetic tilt 
during one solar cycle will mix different polarities (adapted from 
Nerney et al. 1995)}
\label{structure2}
\end{figure}

As the HCS crosses the termination shock, the solar wind velocity drops to $\sim 130km/s$ and keep dropping until when it gets close to the heliopause drops to almost zero, where the flow is deflected to the flanks. The drop is solar wind velocity will make the boundary sectors in the HCS approach each other. Especially near the heliopause the magnetic sectors get tighter and tighter (see Figure \ref{structure2}). We expect that reconnection might play a major role especially in this region, as the magnetic fields of one sector in the heliosheath are pressed against the interstellar magnetic field draped around the heliopause.  For example, considering an upstream solar velocity of 450km/s the separation between the boundary sectors upstream the termination shock, $\lambda$, will be $\sim 7AU$, while immediately downstream the velocity drops by approximately a factor of 3 to 4 to 100km/s so $\lambda$ will drop to 2.0AU. As the velocity approach the Heliopause the velocity drops to zero. Taking a velocity of 10-20km/s the wavelength is reduced to 0.4-0.2AU close to the heliopause (see Figure \ref{view}).

Qualitatively speaking, after the solar magnetic field lines cross the 
termination shock, the spiral gets tighter and tighter due to the 
slow down of the solar wind flow as it approaches the heliopause (see 
Figure \ref{view}). The thickness of the outflow regions in the 
reconnection region depends on the level of turbulence. The length of 
the outflow regions $L$ depends on the mean geometry of magnetic 
field.

In situ measurements in the heliosheath (see Burlaga et al. 2009) 
shows the presence of magnetic sectors beyond the termination shock.
We expect as well the solar cycle to affect the magnetic sectors.
In analytic studies, Nerney et al. (1995) predict a complex region in 
the heliosheath due to the solar cycle effects. Each solar cycle the 
solar global polarity reverses as well. This will create opposite 
polarity sectors due to solar cycle separating strongly mixed 
polarities (due to the tilt of the HCS discussed above). 

Figure \ref{structure2} shows how these two effects will affect the overall 
structure of the heliosheath (adapted from Nerney et al. 1995).
The dark and faint green represents the opposite solar polarity 
sectors while the white represents the regions with strongly mixed 
polarities due to the varying tilt of HCS in each solar cycle. The 
heliopause is represented by a blue line and the termination shock by 
a red line. The heliosheath upstream the termination shock is the far 
right of the figure.
  Their studies, however, were done in the kinematic approximation 
where the magnetic field reaction on the flow was neglected. The flow 
pattern in the heliosheath is expected to be very complex and can 
affect the overall picture above. 

Another effect is related to fact that as the 
solar wind smashes against the interstellar wind the solar magnetic 
field increase in intensity near the solar equatorial plane creating 
magnetic ridges where the interstellar pressure is stronger (see 
Opher et al. 2003, 2004). Thus we expect the solar magnetic field to 
have a very complex structure exhibiting magnetic field reversals 
beyond the termination shock, in the 
heliosheath (see Nerney et al. 1995), as shown in 
Figure~\ref{structure2}.

\section{Model of Reconnection in the Heliosheath}

Currents in resistive plasma drain their energy from magnetic field. However, we shall discuss further that the default rates of magnetic energy conversion that can be obtained from a naive treatment of the problem are too small to be important for the acceleration of energetic particles.

\begin{figure*}[!t]
\includegraphics[width=0.65\textwidth]{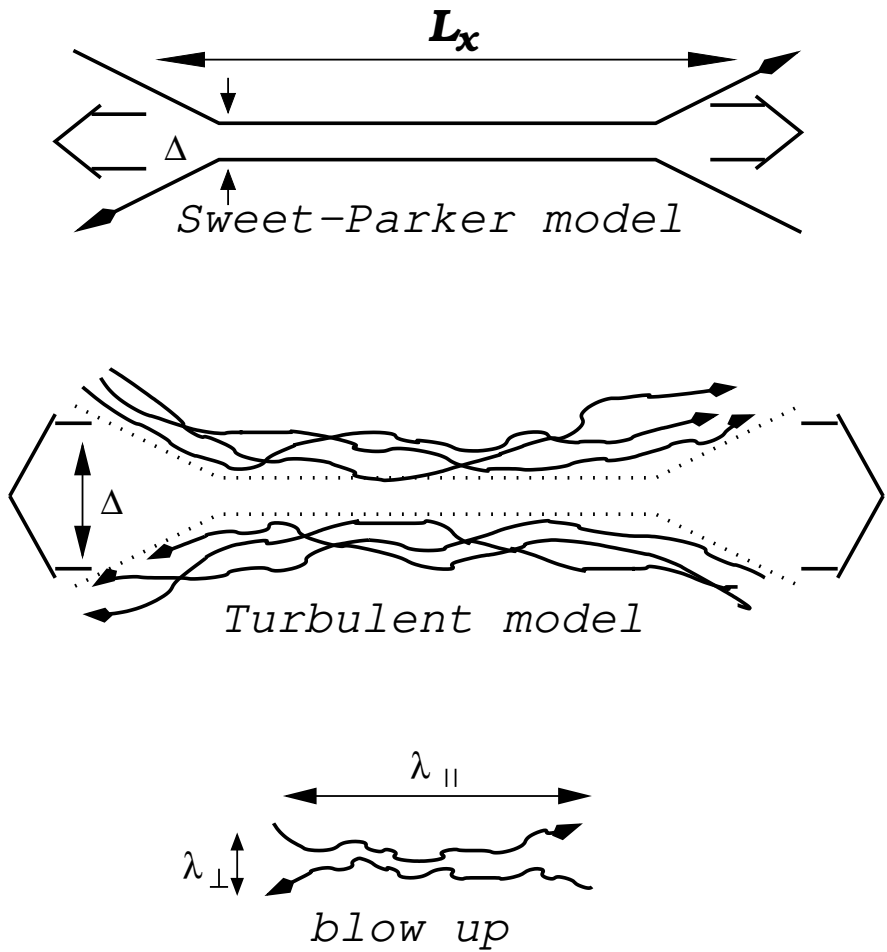}
\caption{{\it Upper plot}: 
Sweet-Parker model of reconnection. The outflow
is limited by a thin slot $\Delta$, which is determined by Ohmic 
diffusivity. The other scale is an astrophysical scale $L\gg \Delta$.
{\it Middle plot}: Reconnection of weakly stochastic magnetic field according to 
LV99. The model that accounts for the stochasticity
of magnetic field lines. The outflow is limited by the diffusion of
magnetic field lines, which depends on field line stochasticity.
{\it Low plot}: An individual small scale reconnection region. The
reconnection over small patches of magnetic field determines the local
reconnection rate. The global reconnection rate is substantially larger
as many independent patches come together. From Lazarian et al. (2004).}
\label{recon1}
\end{figure*}

For instance, the famous Sweet-Parker model of reconnection (Sweet 1958, Parker 1958) (see Figure~\ref{recon1}, upper panel) produces reconnection rates which are smaller than the Alfven velocity by a square root of the
Lundquist number, i.e. by $S^{-1/2}\equiv (LV_A/\eta)^{-1/2}$, where $L$ is the length of a current sheet, $V_A$ is the Alfven velocity, $\eta$ is the Ohmic diffusivity. The current sheet length is determined by transversal extend of the magnetic flux tubes that get into contact and for the heliopause it can be larger than 100 AU. The corresponding reconnection speed for the Sweet-Parker reconnection in this case is negligible being $\sim 3\times 10^{-7}V_A$. This result can be understood intuitively, as in Sweet-Parker reconnection plasma collected over the size $L$ should be ejected with the speed $\sim V_A$ from the thin slot $\delta_{SP}=LS^{-1/2}$, where $S$ for the heliosheath is about $10^{13}$ (see below). The disparity of sizes $L$ and $\delta_{SP}$ makes the reconnection slow and thus one can disregard effect of the Sweet-Parker mechanism for heliospheric reconnection altogether. 

The first model of fast reconnection by Petschek (1964) assumed that magnetic fluxes get into contact not along the astrophysically large scales of $L$, but instead over a scale comparable to the resistive thickness $\delta$, forming a
distinct X-point, where magnetic field lines of the interacting fluxes converge at a sharp point to the reconnection spot. The stability of such a reconnection geometry in astrophysical situations is an open issue. At least for uniform resistivities, this configuration was proven to be unstable and to revert to Sweet-Parker configuration (Biskamp 1986, Uzdensky \& Kulsrud 2000). 

Recent years, have been marked by the progress in understanding some of the key processes of reconnection in astrophysical plasmas. In particular, a substantial progress has been obtained by considering reconnection in the presence of Hall-effect, which is described by the ${\bf J}\times{\bf B}$ term in Ohm's law:
\begin{equation}
{\bf E}+\frac{{\bf v}\times{\bf B}}{c}-\frac{{\bf J}\times {\bf B}}{e n_e c}=\frac{4\pi\eta {\bf J}}{c^2}
\end{equation}
where $e$ is electron charge and $n_e$ is concentration of electrons. Numerical experiments showed that Hall-MHD reconnection is capable of supporting X-points and thus can make the reconnection fast, i.e. comparable to the Alfven speed\footnote{In general, the reconnection is termed fast when the reconnection velocity does not depend on the Lundquist number $S$ or if it depends on $\ln(S)$. In all other cases the large values of $S$ make reconnection too slow for most of astrophysical applications.} (Shay et al. 1998, 2004). 
{\small
\begin{deluxetable}{c|cccc}
\tablecaption{Parameters of heliosheath (from Burlaga et al. 2009). The velocity in the table corresponds to the "bulk velocity" of the flow measured by Voyager 2 near the downstream of the TS.\label{tab:models}}
\tablehead{ \colhead{Name} & \colhead{Magnetic field} & \colhead{Temperature}
& \colhead{Density}& \colhead{Velocity}}
\startdata
Symbol & B & T & n & V\\
Value & 0.1 nT & 1.4$\times 10^5$K & $0.2\times 10^{-3}$cm$^{-3}$ & 150 km/s\\
\enddata
\end{deluxetable}
}

The criterion at which Hall-MHD term gets important for the reconnection is that the ion skin depth $\delta_{ion}$ is comparable with the Sweet-Parker diffusion scale $d_{SP}$. The ion skin depth is a microscopic characteristic and it can be viewed at the gyroradius of an ion moving at the Alfven speed, i.e. $d_{ion}=V_A/\omega_{ci}$, where $\omega_{ci}$ is the cyclotron frequency of an ion. For the parameters of
the heliosheath given in Table~1, we find that for a proton is $d_i\sim 10^3$~km. Thus one can get the constraint on the scale $L$ for which Hall-MHD effects should dominate the reconnection:
\begin{equation}
\frac{\delta_{SP}}{d_{ion}}\approx 0.2 \left(\frac{L}{\lambda_{mfp}}\right)^{1/2}\beta_{pl}^{1/4}<1,
\label{constraint}
\end{equation}
where $\lambda_{mfp}$ is the electron mean free path, where $\beta_{pl}$ is the ratio of thermal pressure to magnetic pressure (see more discussion in Yamada et al. 2006). 

It is clear from the discussion above that the corresponding scales of reconnection which is dominanced of the Hall-MHD term correspond are less than AU, for the parameters in Table~1. We note that the reconnection of with $\frac{\delta_{SP}}{d_{ion}}<1$ is called "collisionless". This may be misleading, as, for instance, a usual definition for being collisionless for magnetized plasma in the interstellar medium (ISM) is to have many Larmor rotations per time between collisions. In this sense the aforementioned requiement for Hall-MHD reconnection is a requirement to be "super-collisionless", as the constraint on the number of collisions is a factor $L/d_{ion}\gg 1$ more stringent that the usually adopted one. Thus magnetic reconnection in most phases of the ISM (see Table~1 in Draine \& Lazarian 1998) is "collisional".  

While the actual scales relevant magnetic reconnection in the heliosheath and near the heliopause are expected to be large (e.g. $L\sim 100$ AU)  it will be clear from the further discussion that this does not actually matter. The Hall-MHD reconnection was demonstrated to provide stable Petschek-type\footnote{Whether this is really Petschek reconnection is still a subject of debates. For instance, slow shocks are, which are an essential feature of Petschek model, are not observed in the simulations (see Zweibel \& Yamada 2009).} configuration characterized by X-points. However, in the presence of external forcing in the heliosheath, e.g. the forcing due to variations of solar wind pressure and turbulence one would expect to observe a collapse of X-points and formation of extended thick outflow regions instead. Incidentally, such regions have been identified
by Ciaravella \& Raymond (2008) via multi-frequency observations of Solar flares. The reconnection within such outflow regions is expected to happen according to a different scheme.

Most astrophysical fluids are turbulent and the heliosheath is not an exception\footnote{If turbulence is very weak, the injection of energy due to reconnection can speed up reconnection, resulting in flares (LV99, Lazarian, Vishniac \& Kowal 2009, Hoang et al. 2009). Thus it is plausible that collisionless Hall-MHD effects can enhance the level of turbulence for the situation when magnetic fields are almost laminar initially. This is a rather unlikely scenario for the healiosheath, however (see more discussion in \S5.2).} with the temporal flows as shown by Richardson et al. (2008). The flows will not only deflect near the heliopause but are expected to be affected near the magnetic ridges that form close to the heliopause (see Opher et al. 2004). In addition, various instabilities are likely to inject turbulent energy. For example, Opher et al. (2003; 2004) suggest the possibility of instabilities near the heliospheric current sheet (HCS), with a narrow jet of high-speed flow, strong wrapping of the HCS, and movement away from the ecliptic. The instability has a characteristic wavelength of tens of AU. Thus the region near the current sheet in the HCS might be unstable and dynamic on the scale of the HCS. Such instabilities could produce high levels of turbulence, back flows, and gradients of density and pressure (see also \S 5.2).

What are the characteristics of magnetic reconnection that we can expect in the heliosheath? We argue in \S5.2 that the characteristics of turbulence may change, especially as a result of flow slowing down and more active reconnection taking place. The characteristic Alfven velocity $V_A$ in the flow downstream the termination shock is $\sim 50$km/s if we use the data in Table~1. Using the John Richardson data from the Voyager 2 flow instrument, of the flows in the heliosheath (which is presented in Opher et al. 2009), we can say that $v_l/V_A \sim 1$. The injection scale of the turbulence $l$ is uncertain. If the flow gets globally turbulent the characteristic size of the flow can provide an estimate of $l$. 

We explain below that the most important for our discussion are the diffusivities and resistivities perpendicular to magnetic field. Thus we adopt the resistivity coefficient $\eta_{\bot}\approx 1.3 \times 10^{13} \ln\Lambda/T^{3/2}$~cm$^2$s$^{-1}$, where $\ln\Lambda\approx 30$ for the parameters in Table~1, and the viscosity coefficient $\nu_{\bot}\approx 1.7\times 10^{-2} n \ln\Lambda/(T^{1/2} B^2)$ (see Spitzer 1962). The rough estimates of the Lundquist and Reynolds numbers ($Re\equiv v_l l/\nu$) are $S\sim 10^{13}$ and $Re\sim 10^{14}$. The exact values of $S$ and $Re$ are irrelevant and the most important message from the above exercise is that the flows are expected to be turbulent (as $Re\gg 1$, see also \S5.2) and the Sweet-Parker reconnection is expected to be negligible (see above).       
 
Turbulence in the helioshath is expected to make magnetic fields at least weakly stochastic. A model of fast 3D reconnection that generalizes Sweet-Parker scheme for the case of weakly stochastic magnetic field was proposed by Lazarian \& Vishniac (1999, henceforth LV99). While the notion of turbulence affecting the reconnection rate was not unprecedented (see Speiser 1970, Strauss 1988, Mathaeus \& Lamkin 1985, 1986), the LV99 model was the first that predicted the change of reconnection rates as the function of turbulence intensity and the turbulence injection scale. It also proved that in the presence of turbulence 3D magnetic reconnection is fast, i.e. independent of $S$, even if only Ohmic resistivities are considered (see also Kowal et al. 2009). 

The middle and lower panels Figure~\ref{recon1} illustrate the key components of LV99 model\footnote{The cartoon in Figure~\ref{recon1} is an idealization of the reconnection process as the actual reconnection region also included reconnected open loops of magnetic field moving oppositely to each other (see the visualization of numerical simulations in Figure~\ref{visual}). Nevertheless the cartoon properly reflects the role of 3 dimentionality of the reconnection process, the importance of small-scale reconnection events and the increase of the outflow region compared to the Sweet-Parker scheme.}. The reconnection events happen on small scales $\lambda_{\|}$ where magnetic field lines get into contact. As the number of independent reconnection events take place simultaneously is $L/\lambda_{\|}\gg 1$ the resulting reconnection speed is not limited by the speed of individual events on the scale $\lambda_{\|}$. Instead, the constraint on the reconnection speed comes from thickness of the outflow reconnection region $\Delta$, which is determined by the magnetic field wandering in a turbulent fluid. The model is intrinsically three dimensional\footnote{Our 2D numerical simulations of turbulent reconnection show that the reconnection is not fast in this case.} as both field wandering and simultaneous entry of many independent field patches, as shown in Figure~\ref{recon1}, are 3D effects. The magnetic reconnection speed gets comparable with $V_A$ when the scale of magnetic field wandering\footnote{Another process that is determined by magnetic field wandering in the diffusion of energetic particles perpendicular to the mean magnetic field. Indeed, the diffusion coefficients of perpendicular diffusion in the Milky Way is just a factor or order unity less than the coefficient of the diffusion parallel to magnetic field (see Jokipii 1999 and ref. therein).} $\Delta$ get comparable with $L$.

For quantitative description of the reconnection one should adopt a model of MHD turbulence (see Iroshnikov 1963, Kraichnan 1965, Dobrowolny, 
Mangeney, \& Veltri 1980,
Shebalin, Matthaeus \& Montgomery 1983, Montgomery \& Turner 1984, Higdon 1984). The most important for magnetic field wandering is the Alfvenic component (see a discussion of the decomposition in \S5.2). Adopting the Goldreich \& Sridhar (1995, henceforth GS95) scaling of Alfvenic component of MHD turbulence extended to include the case of the weak turbulence (see Appendix) LV99 predicted that the reconnection speed in a weakly turbulent magnetic field is
\begin{equation}
V_{R}=V_A (l/L)^{1/2} (v_{l}/V_A)^2
\label{recon}
\end{equation}
where the level of turbulence is parametrized by the injection velocity\footnote{As we discuss in the Appendix, the combination $V_A (v_{l}/V_A)^2$
is the velocity of largest strong turbulence eddies $V_{strong}$, i.e. the velocity at the scale at which the Alfvenic turbulence transfers from the weak to strong regimes. Thus Eq.~(\ref{recon}) can also be rewritten as $V_{R}=V_{strong} (l/L)^{1/2}$.} $v_{l}<V_A$ and the turbulence injection scale $l$. 
 
The scaling predictions given by Eq.~(\ref{recon}) have been tested successfully by 3D MHD numerical simulations in Kowal et al. (2009). This stimulates us to adopt the LV99 model as a starting point for our discussion of magnetic reconnection in heliosphere.

How can $\lambda_{\|}$ be determined? In LV99 model as many as $L^2/\lambda_{\bot}\lambda_{\|}$ events localized reconnection events take place each of which reconnect the flux at the rate $V_{rec, local}/\lambda_{\bot}$, where the velocity of local reconnection events at the scale $\lambda_{\|}$ is $V_{rec, local}$. The individual reconnection events contribute to the global reconnection rate, which in 3D gets a factor of $L/\lambda_{\|}$ larger, i.e.
\begin{equation}
V_{rec,global}\approx L/\lambda_{\|} V_{rec, local}.
\label{global}
\end{equation}

The local reconnection speed, conservatively assuming that the local events are happening at the Sweet-Parker rate can be easily obtained by identifying the local resistive region $\delta_{SP}$ with $\lambda_{\bot}$ associated with $\lambda_{\|}$ and using the relations between $\lambda_{\|}$ and $\lambda{\bot}$ that follow from the MHD turbulence model (see Appendix, Eq.~(\ref{Lambda})). The corresponding calculations in LV99 provided the local reconnection rate $v_l S^{-1/4}$. Substituting this local reconnection rate in Eq.~(\ref{global}) one gets the estimate of the global reconnection speed, {\it if this speed were limited by Ohmic resistivity}, which is larger than $V_A$ by a large factor $S^{1/4}$. As a result, one has to conclude that the reconnection does not depend on resistivity. 

However, it is possible to invert the arguments above and search for the largest scale $\lambda_{\|}$ at which the Sweet-Parker reconnection can reconnect the magnetic field bundles provided the reconnection happens with the speed given by Eq.~(\ref{recon}). Evidently, the reconnection speed at this scale is
$V_{rec,int}\approx V_A\delta_{SP, modified}/\lambda_{\|}$, where $\delta_{SP, modified}$ is a new Sweet-Parker scale related to the reconnection events at the scale $\lambda_{\|}$. Eq.(\ref{global}) is valid for all the scales and not only for the smallest one (Lazarian et al. 2004). Thus substituting there $V_{rec, int}=V_A S^{-1/2} (L/\lambda_{\|})^{1/2}$ and combining the result with Eq.~(\ref{recon}) one gets that
the largest scale Sweet-Parker reconnection events take place at the scale
\begin{equation}
\lambda_{\|}=LS^{-1/3}\left(\frac{L}{l}\right)^{1/3} \left(\frac{V_A}{V_l}\right)^{4/3}
\label{lambdapar}
\end{equation}
and the corresponding thickness of the Sweet-Parker current sheet is
\begin{equation}
\delta_{SP, modified}=L S^{-2/3}\left(\frac{L}{l}\right)^{1/6} \left(\frac{V_A}{V_l}\right)^{2/3}
\label{modified}
\end{equation}

The scale $\lambda_{\|}$ given by Eq.~(\ref{lambdapar}) is much smaller than $L$ and it is evident that for this scale the constraint given by Eq.~(\ref{constraint}) is well satisfied for the heliosphere. Therefore, for all the scales involved in the helispheric reconnection the {\it local} reconnection takes place in the regime when plasma can be considered collisionless.   

What does change if plasma gets collisionless on scales $\lambda_{\|}<<L$ ? As the local reconnection speed does not limit the speed of reconnection in LV99 model, one does not expect the change of the rate given by Eq.~(\ref{recon}). 

The LV99 model is applicable to the situations when plasma effects are included, e.g. Hall-MHD effect, which increases effective resistivities for local reconnection events. While the latter point is difficult to test directly with the existing plasma codes, e.g. with PIC codes, due to the necessity of simulating both plasma microphysics effects as well as macrophysical effects of magnetic turbulence, Kowal et al. (2009) simulated the action of plasma effects by parameterizing them via anomalous resistivities. The values of such resistivities are a steep function of
the separation between the oppositely directed magnetic field lines, which also determines the current separating magnetic fluxes. With anomalous resistivities the structure of the fractal current sheet of the turbulent reconnection changed substantially, but no significant changes of the reconnection rate was reported, which agrees well with the theoretical expectations of the LV99 model. Within the model the explanation of this stems from the fact that the reconnection is already fast (i.e. independent of resistivity) even when small scale reconnection events are mediated by the Ohmic resistivity, while the bottleneck for the reconnection process is provided by magnetic field wandering. Thus the increase of the local reconnection rate does not increase the global reconnection speed. 

As a result, in what follows to describe magnetic reconnection in heliosphere we adopt the LV99 model but with the reconnection events on the scale $\lambda_{\|}$ happening in a collisionless fashion. The latter may have important consequences for the acceleration of the electrons that we discuss below, but is unlikely to effect heavier species.

\section{Magnetic Reconnection and ACR{\small s} acceleration}

\subsection{The mechanism}

Electric field associated with reconnection events can accelerate energetic particles. For a particle of charge $q$ the typical energy gained in such a process is of the order of $q(V_{R}/c) B \lambda_q$, where 
$\lambda_q$ is the coherence length of a particle within the reconnection layer. To accelerate this way one requires to have both  $V_{R}$ and $\lambda_q$ to be large.    
Reconnection as a process of accelerating energetic particles via electric field within the Sweet-Parker reconnection model (see Haswell et al. 1992, Litvinenko 1996) is inefficient due to slow reconnection rates. Moreover, due to the tiny speed of reconnection only a very small fraction of the magnetic energy can be potentially used for driving the acceleration. 

Electric fields are much stronger in the Petschek (1964) reconnection. However, in the Petschek model $\lambda_q$ gets small, which does not allow efficient acceleration either. We may state that in general in any fast reconnection scheme whether this is the Petschek, collisionless, LV99 or any other, the fraction of the volume that is being subject to resistive effects and reveals strong electric fields is small and most of the magnetic energy is converted into kinetic energy. Thus only a small fraction of energy can be transfered through any fast reconnection process to energetic particles if direct electric field acceleration is involved. Therefore we shall ignore this process dealing with ACRs in the heliosphere.

In what follows, for the sake of simplicity, we consider the acceleration of ACRs in an individual reconnection layer. The situations of multiple reconnection layers may present additional effects arising from particles interacting with several layers. However, these situations are not considered in this paper. 

\begin{figure*}
\includegraphics[width=0.45\textwidth]{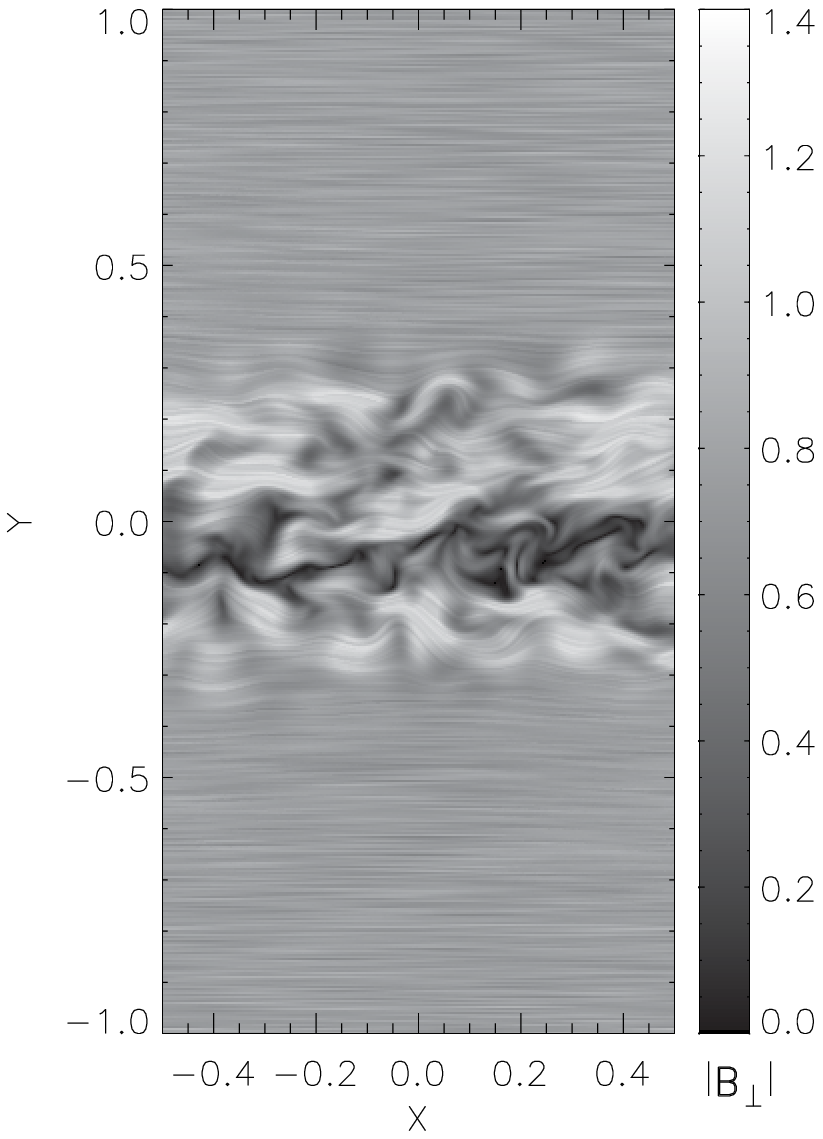}
\includegraphics[width=0.45\textwidth]{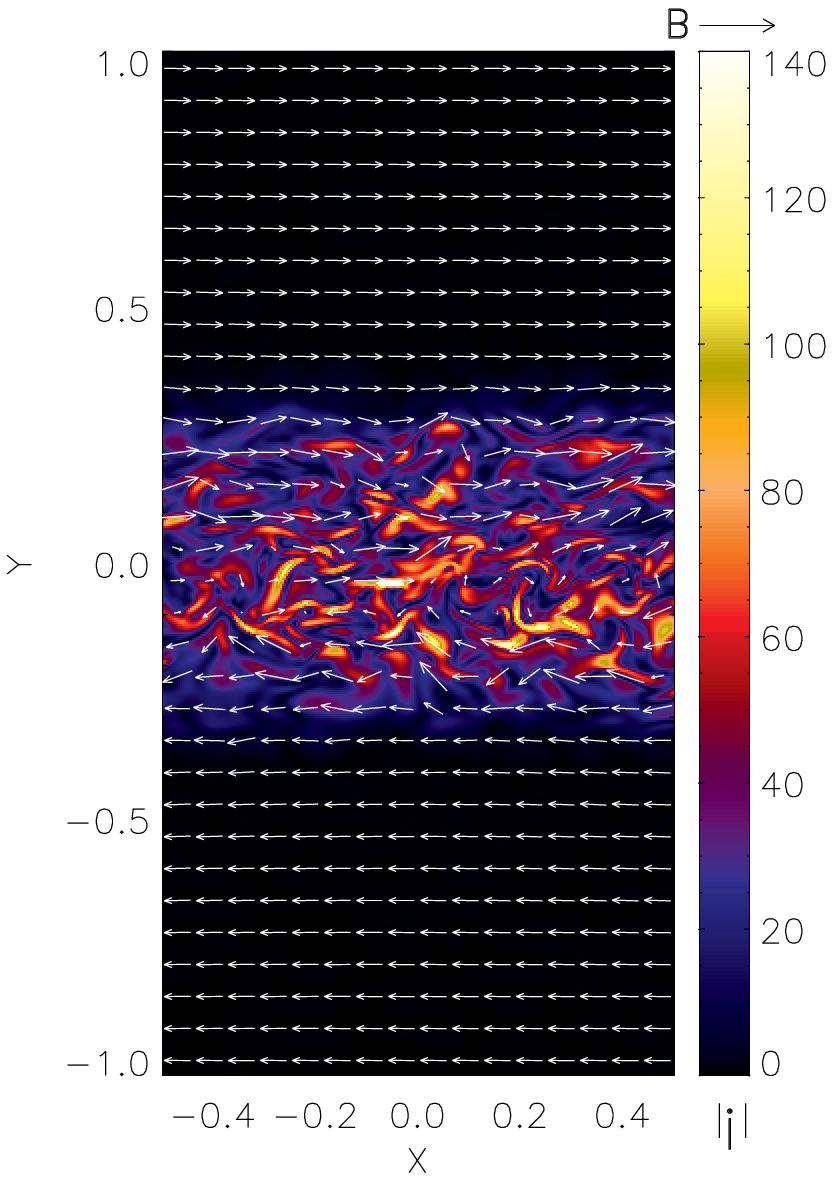}
\caption{Visualization of the 3D ($256\times 256 \times 512$) simulations of reconnection induced by subAlfvenic turbulence in Kowal et al. 2009. The acceleration of energetic particles is expected as the magnetic field lines shrink as a result of magnetic reconnection and particles bounce back and force between the converging magnetic fluxes.
{\it Left Panel}: Topology and strength magnetic field and currents. The structure of magnetic field is visualized by textures (see Burkle, Preuber \& Rumpf 2001). {\it Right Panel}: In the right panel we show distribution of the absolute value of
current density $|\vec{J}|$ overlapped with the magnetic vectors.  The images
show the XY-cut of the computational box where magnetic field reverse their direction. The boxes are elongated in Y-direction to
decrease the influence of the midplane subAlfvenic turbulence driving on the inflow boundaries. \label{visual}}
\end{figure*}

In LV99 scheme the reconnection velocities can approach $V_A$ and therefore be appreciable. Thus one visualizes a different scheme of reconnection, which is similar to the one involving shocks. Indeed we showed that, due to high speeds of stochastic reconnection, particles entrained on reconnecting field lines bounce back and forth between magnetic walls while staying on the field lines that are contracting. This results in a systematic increase of the velocity with every bouncing of energetic particles.
Such a model was discussed in de Gouveia dal Pino \& Lazarian (2003, henceforth GL03, 2005, see also Lazarian 2005), where it was showed that the reconnection induces First order Fermi acceleration of the particles entrained. Figure~\ref{visual} shows the cross-sections of the numerical $256^2\times 512$ box with magnetic field perturbed subAlfvenically, i.e. with the turbulent injection velocity $v_l<V_A$. The field loops of reconnected magnetic flux are clearly visible. Energetic particles are being accelerated as the magnetic field lines of the 3D loops shrink. The effect of shrinking of individual magnetic loops as a result of their complex dynamics in terms of energetic particle acceleration is similar to the particle bouncing back and forth between the upper and lower fluxes (GL03). 
 
\begin{figure}[!t]
\includegraphics[width=\columnwidth]{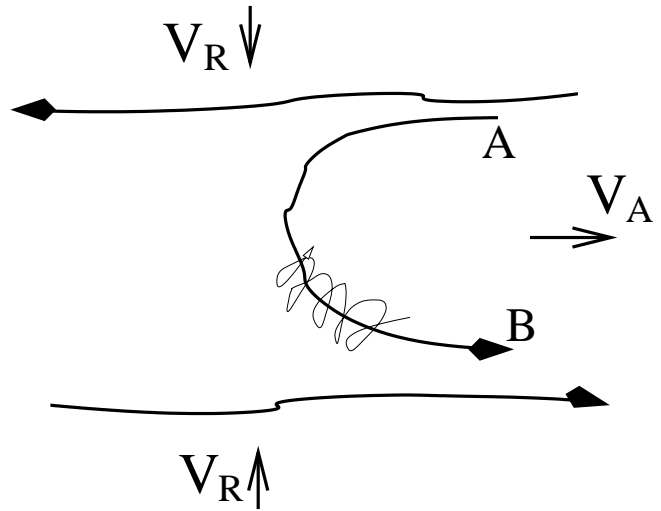}
\caption{  
Cosmic rays spiral about a reconnected magnetic
field line and bounce back at points A and B. The reconnected
regions move towards each other with the reconnection velocity
$V_R$. The advection of cosmic rays entrained on magnetic field
lines happens at the outflow velocity, which is in most cases
of the order of $V_A$. Bouncing at points A and B happens
because either of streaming instability induced by energetic particles or magnetic
turbulence in the
reconnection region. In reality, the outflow region gets filled in by the oppositely moving tubes of reconnected flux (see Figure~\ref{visual}) which collide only to repeat on a smaller scale the pattern of the larger scale reconnection. Thus our cartoon also illustrates the particle acceleration taking place at smaller scales. From Lazarian (2005).}
\label{recon2}
\end{figure}

Figure~\ref{recon2} exemplifies the simplest realization of the acceleration within the reconnection region expected within LV99 model. As a particle bounces back and forth between converging magnetic fluxes, it gains energy. 

The process of acceleration depicted in Figure~\ref{recon2} is easy to quantify. If an energetic particle bouncing back and forth between the magnetic mirrors created by magnetic flux, such a particle having energy $E$ will in every collision gain energy $\sim V_{R}/c E$. The process would continue
till the particle either diffuses perpendicular to the reconnection flux or gets ejected by the outflow together with the plasma and reconnected magnetic flux. The latter possibility was considered in de Gouveia Dal Pino \& Lazarian (2003, henceforth GL03, 2005). Formally, the physical set up considered there corresponds, first of all, to the assumption of the y-component of the diffusion velocity of cosmic rays 
\begin{equation}
V_{y, diff}=\kappa_{yy}/\Delta,
\label{constr}
\end{equation}
 where $\Delta$ is the thickness of the outflow reconnection region and $\kappa_{yy}$ is the diffusion coefficient perpendicular to the average magnetic field, being slow compared to $V_{R}$. This assumption implies that the scattering of energetic particles is efficient and the diffusion coefficient $\kappa_{yy}$ is small. In addition, it assumes
that the diffusion of energetic particles parallel to magnetic field 
\begin{equation}
V_{x, diff}=\kappa_{xx}/L,
\label{constr2}
\end{equation}
where $\kappa_{xx}$ is a coefficient of parallel diffusion, does not exceed $V_A$.

The diffusion coefficient $\kappa_{yy}$ arises from magnetic field wandering as particles diffuse marginally perpendicular to the {\it local} magnetic field. The theory-motivated rates of magnetic field wandering were calculated in LV99 and later used in Lazarian (2006) to understand heat diffusion and in Yan \& Lazarian (2008) to describe cosmic ray propagation. One may argue that the measurements of diffusion parallel and perpendicular to the {\it mean} magnetic field are ill motivated, as the energetic particles do not feel any other field apart from the local field they interact with. 

We believe that all the quantities of the future scattering and acceleration theories should be formulated in terms of local fields, the same way as the turbulent theories are formulated (see Appendix). If, nevertheless, one adopts a conventional approach, for the developed strong turbulence at scales $\Delta$ less than the turbulence injection scale $l$ the perpendicular diffusion coefficient for energetic particles is $\kappa_{yy}\approx (\Delta/L) (v_l/V_A)^4 \kappa_{xx}$ (Yan \& Lazarian 2008). Substituting the latter expression in Eqs.~(\ref{constr}) and comparing $V_{y, diff}$ with $V_R$ given by Eq.~(\ref{recon}), one may notice that for subAlfvenic turbulence with $l\sim L$ this constraint is less restrictive that the constraint arising from  $V_{x, diff}<V_A$. For the sake of simplicity in what follows we shall refer to $V_{diff}$ as the most restrictive of the two processes. 

If one uses the experimentally measured parallel diffusion coefficient suggested by the Palmer consensus (Palmer 1982), namely, the diffusion coefficient corresponding to the parallel mean free path for particles between 0.5Mev and 5Gev being from 0.08AU to 0.3AU, $V_{diff}$ given gets larger than the corresponding velocity of advection of cosmic rays with reconnected magnetic field. However, for the diffusion along magnetic field lines undergoing reconnection the aforementioned estimate for the diffusion speed is definitely an overestimate, as both magnetic reconnection and streaming particles should create magnetic perturbations that efficiently scatter particles back (see Cesarsky 1980). The former effect of modification of turbulence in turbulent reconnection was predicted in LV99 and observed in simulations by Kowal et al. (2009)\footnote{The complex interaction of modes generated through streaming instability and ambient turbulence was addressed in Yan \& Lazarian (2002), Farmer \& Goldreich (2004), Lazarian \& Beresnyak (2006) and Beresnyak \& Lazarian (2008b). However, the treatment is limited to Alfvenic modes only, and it does not include reflection of Alfvenic modes from the inhomogeneities in density and magnetic field. Thus it is not directly applicable to the turbulence strongly affected by hierarchy of reconnection events as is the case for LV99 model.}. Thus we expect that the particles to be both reflected back by magnetic bottles and efficiently scattered by the modified turbulence.            

To derive the energy spectrum of particles one can use the routine way of dealing with first order Fermi acceleration (see Longair 1992). Consider the process of acceleration of $M_0$ particles with the initial energy $E_0$. If a particle gets energy $\beta E_0$ after a collision, its energy after $m$ collisions is $\beta^m E_0$. At the same time if the probability of a particle to remain within the accelerating region is $P$, after $m$ collisions the number of particles gets $P^m M_0$. Thus 
$\ln (M/M_0)/\ln(E/E_0)=\ln P/\ln\beta$ and
\begin{equation}
\frac{M}{M_0}=\left(\frac{E}{E_0}\right)^{\ln P/\ln\beta}
\end{equation}
For the stationary state of accelerated particles the number $M$ is the number of particles having energy equal or larger than $E$, as some of these particles are not lost and are accelerated further. Therefore:
\begin{equation}
N(E)dE=const\times E^{-1+(\ln P/\ln\beta)} dE
\label{NE}
\end{equation}

To determine $P$ and $\beta$ consider the following process. The particles from the upper reconnection region see the lower reconnection region moving toward them with the velocity $2V_{R}$ (see Figure~\ref{recon2}). If a particle from the upper region enters at an angle $\theta$ into the lower region the expected energy gain
of the particle is $\delta E/E=2V_{R}\cos\theta/c$. For isotropic distribution of particles their probability function is $p(\theta)=2\sin\theta\cos\theta d\theta$ and therefore the average energy gain per crossing of the reconnection region is
\begin{equation}
\langle \delta E/E \rangle =\frac{V_{R}}{c}\int^{\pi/2}_{0} 2\cos^2\theta \sin\theta d\theta=4/3\frac{V_{R}}{c}
\end{equation}
An acceleration cycle is when the particles return back to the upper reconnection region. Being in the lower reconnection region the particles see the upper reconnection region moving the speed $V_{R}$. As a result, the reconnection cycle provides the energy increase
$\langle \delta E/E \rangle_{cycle}=8/3(V_{R}/c)$ and
\begin{equation}
\beta=E/E_0=1+8/3(V_{R}/c)
\label{beta}
\end{equation}  

Consider the case of $V_{diff}\ll V_R$. The total number of particles crossing the boundaries of the upper and lower fluxes is $2\times 1/4 (n c)$, where $n$ is the number density of particles. With our assumption that the particles are advected out of the reconnection region with the magnetized plasma outflow the loss of the energetic particles is $2\times V_{R}n$. Therefore the fraction of energetic particles lost in a cycle is $V_{R} n/[1/4(nc)]=4V_{R}/c$ and
\begin{equation}
P=1-4V_{R}/c.
\label{P}
\end{equation} 

Combining Eq.~(\ref{NE}), (\ref{beta}), (\ref{P}) one gets
\begin{equation}
N(E)dE=const_1 E^{-5/2}dE,
\label{-5/2}
\end{equation}
which is the spectrum of accelerated energetic particles for the case when the back-reaction is negligible (see also GL03)\footnote{The obtained spectral index is similar to the one of Galactic cosmic rays.}.

We note that Eq.~(\ref{-5/2}) provides the estimate of the expected spectrum for a rather idealized situation. It is important to understand what modifications of the spectrum we expect to observe in more realistic circumstances. First of all, the derivation above considers only particles bouncing between from lower and upper reconnecting fluxes. The actual picture of the stochastic reconnection in LV99 includes many reconnection events happening at different scales (see Figure~\ref{visual}). However, each of these events can be viewed as a repetition of the large-scale reconnection event and should provide the same type of power spectrum. In fact, the efficiency of acceleration increases with the decrease of the scale of reconnection (see \S 4.3).

Dealing with the reconnection, we have another limiting case which can be easily treated. In the case when $V_{x, diff}\ll V_{y, diff}$ and $V_{y, diff}\gg V_R$ one can formally take $P\approx 1$, as the ejection of the particles is negligible. This situation corresponds to an infinite reconnection region where particles may diffuse into magnetic flux, but do not escape. In this situation Eq.~(\ref{NE}) provides 
\begin{equation}
N(E)dE=const_2 E^{-1}dE
\label{-1}
\end{equation}
which coincides with the spectrum obtained in Jokipii (2009) by solving one dimensional Parker equation for the Sweet-Parker reconnection process. Naturally, the model of particle acceleration which does not  allow for particle escape is rather artificial. However, our discussion of this situation exemplifies the fact that the decrease of the particle escape makes the spectrum of accelerated particles more shallow. Similarly, one can show that the diffusion along magnetic field lines in the $x$-direction, that enhances particle escape should make the spectrum of the energetic particles steeper.

\subsection{Particle backreaction}

The backreaction of accelerated particles is known to be important for the processes of shock acceleration, although the consensus on the quantitative description of the process is still missing (see Malkov \& Diamond 2009). Thus it is not surprising that the situation with the backreaction of particles in the reconnection sites is rather unclear. The only work attempting to address this problem that we are aware of is that by Drake et al. (2006, henceforth DX06). DX06 repeats the claim in GL03 that the fist order Fermi acceleration should happen as a result of interaction of energetic particles with reconnecting magnetic fields. However the study is intended for the  acceleration of electrons and the model is of 2D collisionless reconnection. In this scheme of reconnection the production of contracting magnetic loops is expected. Within these loops the energetic particles are expected to undergo acceleration. 

The acceleration of particles other than electrons was not discussed by DX06 within their model. For the small scale loops considered by DX06 the Hall term dynamics beyond the ion skin depth is important. In both cases the contracting magnetic loops increase the energy of entrained particles. In fact, the evidence for such process can be seen in the simulations of test particles in the magnetotail (Birn, Thomsen \& Hesse 2004) and traced back even further to the test particles studies in MHD models with magnetic islands  (Matthaeus, Ambrosiano \& Goldstein 1984, Kilem 1994).

DX06 appeals to physical arguments rather than to direct numerical calculations to justify the advocated picture of backreaction. In particular, the backreaction is introduced by the term $(1-8\pi{\bar\epsilon}_{\|}/B^2)$, where $\bar{\epsilon_{\|}}$ is parallel energy of energetic particles averaged over the distribution of particle velocities. This term gets negative to represent the halt in the contraction of the magnetic loops. The generalization of this picture for open loops expected for a generic 3D reconnection is feasible, but was not performed by DX06. 

We feel that DX06 model, although formulated in terms of contracting 2D loops, can potentially be generalized for contracting 3D spirals, in which case the acceleration processes in GL03 and DX06 are
similar\footnote{We feel that the potential deficiency of considering closed loops compared to the open loops in the acceleration picture advocated in this paper is that in the incompressible limit and in the presence of particle scattering we expect to see the cancellation of the increase of the total particle momentum, as the increase of the particle momentum $p_{\|}$ through magnetic loop contraction considered by DX06 and the decrease of the momentum perpendicular to the magnetic field $p_{\bot}$ through the betatron effect, as it was shown for incompressible MHD turbulence in Cho \& Lazarian (2006). In the absence of the collisions the increase of the parallel momentum for a contracting loop is limited by the relation on the particle energy $E$ and the loop length $l_{loop}$ of the form $El_{loop}^2=const$. It is easy to see that the escape of particles from one loop and capturing into another should result in random changes in the projections of particle momentum and the effect of this loop changing is similar to collisions, which makes the limitations on acceleration derived in Cho \& Lazarian (2006) applicable again. These limitations are not applicable to the case of the open loops, which ends are connected with converging fluxes that we discussed above (see also Somov \& Kosugi 1997, Giuliani, Neukirch \& Wood 2005).}. 
The notable difference is that with Hall-MHD effect included the loops can preferentially accelerate electrons on the scales less than the ion skin depth $d_{ion}$. 

In terms of the nature of the particle backreaction electrons and protons should act similarly as soon as their energy gets comparable with the energy of the magnetic field that drives them. This, according to DX06 may result in a more shallow spectrum, with the index $Ð3/2$ rather than $-5/2$ predicted by Eq.~(\ref{-5/2}). 

It is not clear to what extend the backreaction is important for the energetic particles accelerated from low energy (around keVs) into the MeVs ACRs (from 0.04-4MeV) in the zones of reconnection in the heliosheath. Thus, depending on the importance of the yet unclear backreaction of the energetic particles, the actual spectral index may vary from $-5/2$ to $Ð3/2$, which encompasses the value of $Ð5/3$ that was observed by the Voyagers. The fact that the observed value is close to the low boundary of the expected indexes means that the backreaction of anomalous cosmic rays to magnetic loops contracting in the process of reconnection is probably important.  

Contracting loops are not a part of the picture of reconnection in the limits $V_{y,diff}>V_R$ or/and $V_{x,diff}>V_A$. For these cases, backreaction of the particles may be very different. We expect both parallel diffusion to increase and the perpendicular diffusion to decrease with the increase of the strength of the shared component of magnetic field. We also expect that for higher energy particles the confinement within the reconnection region gets problematic, which limits the energies to which the energetic particles can be accelerated. For instance, the upper boundary on the energies of the accelerated energetic particles comes from the requirement that the particle Larmor radius $R_L$ is less that the transverse dimension of the magnetic flux undergoing reconnection. Assuming that the latter is about 1AU, we get the maximal energy of particles the magnetic field of $0.1$~nT (see Table~1) of the order of $10^3$~Mev, which is higher than the energies of the ACRs. However, as the particles approach this energy their diffusion and escape in the $x$-direction to increase, changing the nature of reconnection and decreasing its efficiency. Needless to say that these and similar issues require a further quantitative investigation. 

\subsection{Maximal rates of acceleration}

The acceleration efficiencies for ACRs are estimated to be high with the acceleration time for roughly 100 MeV ACR is being less than a year (Mewaldt et al. 1996). This induces constraints on the possible acceleration mechanism responsible for the origin of these particles. Appealing to an analogy of the first order Fermi acceleration at reconnection sites and magnetized shocks one can use the relevant results in Jokipii (1992). 

Indeed, the most efficient acceleration of ACRs is expected when they experience bouncing of the converging mirrors every Larmor period. Such mirrors are expected to be created by small scale reconnection events arising from oppositely moving flux tubes within the reconnection outflow region. Such events are predicted in LV99 and can be seen in the visualizations of the reconnection regions in the presence of turbulence (see Figure~\ref{visual}). The particle within these microscale reconnection regions behave analogously to a particle accelerated in a perpendicular shock considered in Jokipii (1992). For a particle of the Larmor radius $R_L$ one can roughly estimate the corresponding acceleration time as $\tau_{accel}>2R_L/V_A$. The latter estimate is a factor of unity different from the estimate obtained by Jokipii (1992) for a perpendicular shock where $8R_L/V_{shock}$ was obtained. Thus the estimates of a minimal acceleration time, which are less than a month in Jokipii (1992), are applicable to the process we consider in the paper.

\section{Discussion and Summary}

This paper is of exploratory nature. We propose an alternative mechanism for explaining the origin of ACRs appealing to fast magnetic reconnection. Some time ago, such an appeal would sound completely speculative due to the notoriously enigmatic nature of magnetic reconnection. However, as a result of recent progress in understanding of magnetic reconnection, we believe that the time is right to start discussing the consequences of the process.

\subsection{Expected sites of acceleration}

Current sheets are common in the heliosphere. However, not all current sheet are expected to be associated with particle acceleration. For tangible particle acceleration, the reconnection rates should be high, i.e. comparable with $V_A$. In this respect, LV99 model predicts that  the reconnection rate may be low or high depending on the level of turbulence (see Eq.~(\ref{recon}))\footnote{ An additional dependence comes from the boundary conditions. If plasma and shared magnetic flux, i.e. the magnetic flux associated with the guide field, are constrained from freely leaving the reconnection zone, the reconnection rates decrease compared to those given by Eq.~(\ref{recon}).}. 

When the reconnection velocity is a small fraction of the solar wind speed (e.g. as this is the case prior to the termination shock shown in Figure~\ref{structure1}), then the reconnection is expected to play a marginal role in the particle acceleration. However, as the flow slows down after the termination shock, magnetic field reversals are coming closer and get crowded, which is illustrated in Figure~\ref{structure1} by current sheets crowding after the termination shock. As this happens we expect a larger portion of magnetic flux to be consumed by reconnection per unit time and a larger density of the accelerated energetic particles. 

In addition, the large scale variations of magnetic field associated with solar cycles are expected to contribute to the acceleration of energetic particles. It is clear from Figure~\ref{structure2} that the corresponding effect is negligible within the termination shock. In fact, "touching" X-point Petschek-type events are expected before the termination shock (see Gosling et al. 2006a). However, it is also clear that magnetic fields associated with different cycles are pressed together as we approach to the heliopause. This should shift the sites of the energetic particle acceleration towards the heliopause.

We should mention that the direct in situ studies of reconnection (see Gosling et al. 2005a, Gosling et al. 2007, Phan, Gosling \& Davis 2009) have not revealed the excess of the energetic particles associated with magnetic reconnection exhausts in the solar wind (Gosling et al. 2005b). We believe that this can be due to the fact that the X-point Petschek-type reconnection, signatures of which were searched for, is inefficient in accelerating energetic particles as we discussed in \S4. In the paper we appeal to thick turbulent reconnection regions (see Fig~\ref{recon1}) as the sites of acceleration\footnote{Alternatively, in the 2D picture employed in Drake et al. (2006) the loops within the thick reconnection regions rather than X-points are the engines of the accelerated particles.} We expect such regions to emerge as magnetic field reversals get crowded and magnetic field lines of different direction press against each other making the X-type opening of reconnection regions problematic (see the discussion in \S 2).

Finally, let us mention the current sheets associated with magnetic turbulence in the absence of large scale magnetic field reversals. The variations of magnetic field directions induced by turbulent motions induce their own small scale current sheets. However, for subAlfvenic driving the variations of the magnetic field induced by turbulence are not large. Only the oppositely directed components of magnetic field are involved in the reconnection process and that it is they that determine the Alfven speed that enters Eq.~(\ref{recon}). Thus the large scale reversals are much more preferable sites for energetic particle acceleration. However, the first order Fermi acceleration within reconnection layers created by MHD turbulence may not always be neglected, especially when compared to the second order Fermi acceleration by turbulence (see \S5.5).  

\subsection{Parameters of turbulence}

According to Eq.~(\ref{recon}) the rate of reconnection depends on the level of turbulence, i.e. on $v_l$, and the injection scale of the turbulence $l$. The two parameters are still uncertain in depth of the heliosheath towards the heliopause, where we expect most of the magnetic reconnection events to take place. However, even very weak turbulent driving corresponding to a small fraction of $V_A$ provides substantial reconnection rates incomparably larger than the Sweet-Parker prediction (see \ref{recon}). The dependence on $l$ is rather weak in Eq.~({\ref{recon}) and therefore an estimate $l\sim L_x$ should be acceptable. 

Why are we sure of the presence of turbulence at the distances which have not been probed by Voyagers so far? One may claim on very general grounds that astrophysical turbulence is the consequence of high Reynolds numbers of astrophysical flows (see a discussion in Lazarian et al. 2009). Generically, hydrodynamic flows get turbulent for $Re\sim 10$ or 100. A notable exception of this rule are Keplerian flows in accretion disks, which, however, get also turbulent for large $Re$ in the presence of weak magnetic fields (Velikhov 1959, Chadrasekhar 1960, Balbus \& Hawley 1991). For the magnetized flows in the heliosheath the damping of Alfvenic perturbations is limited by the perpendicular viscosity discussed in \S3. Thus the expected $Re$ will be larger than $10^{10}$, which much exceeds the threshold for the fluid getting turbulent. 

Recent numerical simulations support theoretical conclusions that magnetic reconnection can be self-regulating process and the initial level of magnetic stochasticity may matter little as the reconnection proceeds. For instance, 3D MHD simulations in Hoang et al. (2009) showed that magnetic fluxes of different directions in low-$\beta$ plasma brought in the contact at $t=0$ develop turbulence and dissipate the flux within several crossing Alfvenic times (see also Bettarini \& Lapenta 2009). 

It is very advantageous that the reconnection in LV99 depends only on the Alfvenic component of MHD turbulence, as the compressible components are subject to more damping (see Brunetti \& Lazarian 2007) and also more fancy channels of cascading (see Chandran 2005). The possibility of segregating of Alfvenic component from the rest of the MHD cascade is suggested theoretically in GS95 and proven in direct 3D compressible MHD numerical simulations (Cho \& Lazarian 2002, 2003). In the fully ionized plasma  of the heliosheath the Alfvenic component cascades to the proton Larmor radius (see the discussion of the cascade in a partially ionized gas in Lazarian, Vishniac \& Cho 2004). 

We discuss in the Appendix A that the possible uncertainties of the spectral index of the Alfvenic turbulence that are consistent with the observations, in situ spacecraft measurements and numerical simulations do not change the LV99 conclusion that the reconnection is fast. Thus the model suggested in the paper does not depend on the outcome of the ongoing debates on the exact spectral index of Alfvenic turbulence. Similarly,  in the presence of the backreaction of accelerated particles the turbulence spectrum is expected to be modified, but the rates of reconnection and therefore rates of ACRs accelerations are not expected to change appreciably.

\subsection{Predictions and limitations of the model}

In the paper above we proposed that the acceleration of particles arising from magnetic reconnection could explain the origin of the ACRs. In particular we predict  that the source of the ACRs  can be deeper in the heliosheath, close to the heliopause. We maintain that the acceleration of energetic particles by magnetic reconnection is an unavoidable process. Our idea can be tested as the ACR energy spectra of H, He, N, Ne and O in the heliosheath will slowly unroll as Voyager 1 and 2 make their way into the heliosheath.

The actual predictions of Voyager 1 and 2 measurements require detailed numerical modeling of the energetic particle propagation in the heliosheath, which is a problem far from being handled reliably. Qualitatively speaking, we expect to see the variations of the anisotropy of the ACRs as Voyagers approach and cross reconnection regions. The turbulent reconnection regions are characterized by substantial changes in the direction of magnetic field, presence of magnetic flux at substantial angles to the magnetic field in the adjacent regions and magnetic field reversals of this unusually aligned flux (see Figure~\ref{visual}). We expect also to see an increased level of velocity fluctuations as the reconnection itself drives turbulence. At the same time, the events at scales less than $\lambda_{\|}$ given by Eq.~(\ref{lambdapar}) can show signatures of collisionless reconnection. Detailed predictions concerning ACRs will get available as the theory of acceleration in turbulent reconnection region matures (see the discussion of problems below).  

Difficulties associated with understanding of astrophysical acceleration processes are well known. Even for the well-recognized process of pariticle acceleration in shocks (Axford, Leer \& Skadron 1977, Krymsky 1977,
Bell 1978, Blandford \& Ostriker 1978) the details of acceleration are unclear (see Malkov \& Diamond 2009). It is not surprising that for the first order Fermi acceleration within a turbulent reconnection zone we are groping for basic facts. Obviously, the model we presented in \S4.1 is simplistic. It does not prescribe the energetic particle diffusion coefficients, does not account for the compressibility of the medium, disregards of the possibility of the interaction between the adjacent reconnection sites etc. Thus, it is natural that the uncertainties of the expected particle spectrum that we get are much higher and we have to be satisfied with rough consistency of the theoretical expectations to the observed spectra of ACRs. Below we discuss the potential uncertainties of the models. We expect that future numerical simulations will resolve the outstanding issues that we outline.    

The efficiency of particle acceleration in fast reconnection has not been resolved yet. It is clear, that the competition will be between the channeling of magnetic energy into the kinetic energy of thermal plasma and energetic particles. The solar flare reconnection events can provide us with a hint. Indeed, the observations SME indicate that most of the energy is going into energetic particles. Similarly, the analogy between the shock acceleration and the first order Fermi acceleration in the reconnection regions is also suggestive of higher percentage of energy going into the acceleration of energetic particles. Future research should clarify this issue. The self-consistent model of the generation of the ACRs in the heliosheath should include the reconnection layers acting as a source term for solving the transport equation for energetic particles (see Schlickeiser 2003) . Unfortunately, with the complex and yet unknown structure of the heliosheath it may be currently impossible to reliably model the energetic particle injection and propagation. 

The fact that magnetic reconnection may be strongly affected by the backreaction of the energetic particles does not necessarily ensure that the magnetic field and energetic particles get into equipartition throughtout the entire heliosheath region. Indeed, particles may escape efficiently with the reconnected flux and the parts of healiosheath not subject to reconnection may have the energy density of particles well below the equipartition. At the same time, the accumulation of particles above the equipartition value is prohibited due to insufficiency of magnetic tension for constraining them. If the energy of energetic particles gets locally larger than the magnetic energy one expects to observe local expanding magnetic bubbles, which will burst creating open magnetic field lines (similar to the open magnetic field lines of Solar wind) unless constrained by the pressure of the ambient regions.    

In our calculations in \S4 we assumed that energetic particles entering the unreconnected flux (e.g. upper flux tube in Figure~\ref{recon2}) get scattered and randomized before they leave it to collide with the oppositely moving magnetic flux (e.g. lower flux tube in Figure~\ref{recon2}). This may be not the only process involved. For instance, energetic particles can be reflected by magnetic mirrors, which will not change the adiabatic invariant of the particle. In this case, only parallel component of the particle momentum will grow. However, as it was argued in Lazarian \& Beresnyak (2006) such a distribution is unstable in respect to gyroresonance instability. This instability is known to generate circularly polarized Alfvenic perturbations with ${\bf k}\|{\bf B}$ and those will scatter and randomize particles, thus decreasing the anisotropy. Therefore we do not expect a change of the acceleration efficiency.           

 At the same time, the damping of turbulence at the Larmor radius scale of protons makes the acceleration of minor ions preferable. Indeed, scattering of energetic particles is an important component of the acceleration process. This scattering depends on the presence of magnetic turbulence at the Larmor radius scales. As this scale is larger for the minority ions, their acceleration may proceed more efficiently, which is also the case for the acceleration in shocks (see Mewaldt et al. 1996). 

\subsection{Role of reconnection microphysics}

Let us show how the assumed microphysics of reconnection affects the energetic particle acceleration. 
As we discussed in \S3, the adopted LV99 model of magnetic reconnection in the heliosheath includes small scale event mediated by collionless effects\footnote{It is interesting to note that some of the the in situ measurements of the parameters of the current sheet may be consistent with the predictions of the collisionless reconnection, while the actual reconnection is going according the LV99 scheme.}  and large-scale reconnection limited by magnetic field wandering according to the LV99 model. While in \S3 we argue that collisionless effects are irrelevant in terms of overall rates of reconnection, their presence may be important for the acceleration of electrons. One might also expect that the dynamics of the magnetic loops at scales smaller than the proton gyroradius may be different from the magnetic loops on larger scales as a result of the difference in the properties of whistler (electron MHD) and Alfvenic turbulence (Cho \& Lazarian 2004, 2009).

For our discussion of the acceleration in a reconnection layer we adopted the collisionless model of reconnection on the small scales $\lambda_{\|}$. A number of issues related to the reconnection in this regime are still a subject of debates. For instance, the role of the Hall term is challenged in a number of papers (see Karimabadi et al. 2004, Bessho \& Bhattacharjee 2005, Daughton \& Karimabadi 2007). Whether the collisionless reconnection is fast has been also questioned (see Wang et al. 2001, Fitzpatrick 2004, Smith et al. 2004). In addition, small-scale events may proceed at different rates due to the external forcing preventing formation of small-scale X-points within a turbulent reconnection region. 

We believe that while some details of the acceleration may change, e.g. the acceleration of electrons may be modified in the absence of the Hall term, the major elements of the model of the acceleration of energetic particles in heliosheath, that we discuss in the paper will stay. First of all, even if collisionless reconnection is not fast, this should not change the overall speed of the reconnection of the weakly turbulent magnetic field (LV99). Then, the contracting open magnetic loops accelerating electrons are still likely to emerge at scales smaller than proton Larmor radius due to the turbulent motions protruding at the electron MHD scale. Future numerical calculations should clarify these issues.

\subsection{Role of turbulent acceleration of ACRs}

While the alternative idea of blunt shocks stays a viable one, we have doubts about the idea of second-order Fermi acceleration of energetic particles observed by Voyagers. First of all, the estimates in Jokipii (1992) suggest that the time for the second order Fermi acceleration by turbulence is prohibitively long. Then, below we claim
that the interacting large-scale turbulent eddies can produce reconnection regions inducing the first order Fermi acceleration, which sometimes may be more efficient than the traditional second order Fermi acceleration.  

It is well known that apart from accelerating the energetic particles in the heliopheric current sheets, magnetic reconnection can induce turbulence that can accelerate energetic particles (see Petrosian et al. 2006 and ref. therein). Indeed, in any scheme of fast reconnection only small fraction of energy is consumed through direct plasma heating. While further research is necessary at this point, it is reasonable to assume that the partition of energy that is released directly in the reconnection zone and is available in the form of magnetic tension after magnetic field lines leave the reconnection zone depends on the initial configuration of the reconnecting magnetic fluxes. If most of the energy is accumulated in the form of magnetic field lines bended outside the reconnection region, as this is the case in some models of solar flares (see Tsuneta 1996), then turbulence may absorb most energy released in the event. Although we do not believe that this is the case for the magnetic fields in the heliosheath, even in this unlikely hypothetical situation we expect to see bursts of turbulent reconnection accelerating energetic particles. 

Interestingly enough, we may argue that the interaction of energetic particles with reconnection regions naturally arising in magnetic turbulence may result in the first order Fermi acceleration.    
It is easy to see that reconnection events in MHD turbulence should happen through every eddy turnover (see LV99). For small scales magnetic field lines are nearly parallel (see Appendix) and, when they intersect, the pressure gradient is not $V_A^2/\lambda_{\|}$ but rather $(\lambda_{\bot}^2/\lambda_{\|}^3) V_A^2$, since only the energy of the component of the magnetic field that is not shared is available to drive the outflow. On the other hand, the characteristic length contraction of a given field line due to reconnection between adjacent eddies is $\lambda_{\bot}^2/\lambda_{\|}$. This gives an effective ejection rate of $V_A/\lambda_{\|}$. Since the width of the diffusion layer over the length $\lambda_{\|}$ is $\lambda_{\bot}$ the Eq.(\ref{recon}) should be replaced by $V_{R}\approx V_A (\lambda_{\bot}/\lambda_{\|})$, which provides the reconnection rate $V_A/\lambda_{\|}$, which is just the nonlinear cascade rate on the scale $\lambda_{\|}$. This ensures self-consistency of the critical balance for strong Alfvenic turbulence in highly conducting fluids (LV99). Indeed, if not for fast turbulent reconnection the buildup of unresolved magnetic knots is unavoidable, flattening the turbulence spectrum compared to the theoretical predictions. The latter contradicts both to Solar wind measurements and to numerical calculations. 

The energy of reconnected magnetic field during the eddy turnover is comparable with the energy of the eddy. In the absence of cosmic ray acceleration the energy liberated in reconnection goes into motions comparable to the dimensions of the reconnecting eddies, so this energy release will not short circuit the turbulent energy cascade. In the presence of cosmic ray acceleration a substantial part of the turbulent energy may go into the energetic particle acceleration\footnote{R. Jokipii (1999) points out that the coexistence of two big power laws in the sky: the one by cosmic rays and one by turbulence shows the fundamental inter-relation of turbulence and cosmic rays. One may speculate that the process of first order Fermi acceleration in the turbulent eddies may be at the core of this inter-relation.} and therefore the second order Fermi acceleration in turbulence may play a subdominant role even for the part of the magnetic energy that was released beyond the reconnection region and induced magnetic turbulence.     

Consider the problem from the point of view of energetics. For driving with turbulent injection velocity
of the order of Alfven velocity, around 80\% of energy go into Alfvenic incompressible modes (Cho \& Lazarian 2003). This result obtained with solenoidal driving of turbulence corresponds well to the results for hydrodynamic turbulence where, for arbitrary driving, a substantial part of energy goes into solenoidal motions (Biskamp 2003, see also Federath et al. 2009 for the MHD case). At the same time, Alfvenic modes arising from large scale driving are shown to be
very inefficient for accelerating energetic particles due to the high anisotropy of Alfvenic modes (see a discussion Cho \& Lazarian 2006). The acceleration by fast modes (Yan \& Lazarian 2002, 2004, 2008, Brunetti \& Lazarian 2007) is limited\footnote{Here we consider resonant and Transient Time Damping (TTD) acceleration
(see Schlickeiser 2003) and disregard the acceleration arising due to large scale contractions as the study in Cho \& Lazarian (2006) showed that this acceleration process is subdominant unless the turbulence driving is superAlfvenic. The latter is not expected for the turbulence arising from magnetic reconnection. In our discussion we also disregarded the process of generation of waves with $k$-vectors parallel to magnetic field ("slab modes") through gyroresonance instability as predicted in Lazarian \& Beresnyak (2006).}  by fast modes collisionless damping. This can be viewed as an additional argument against second order Fermi acceleration of ACRs. 

All in all, while turbulence is essential for driving magnetic reconnection, the role of turbulence in the second order Fermi acceleration of ACRs may be subdominant. We argue that turbulence itself may be associated with first order Fermi acceleration related to the local reconnection of the magnetic field of adjacent eddies. If proven by further research, this may substantially increase the potential of turbulence in accelerating energetic particles in the situations beyond the one we considered above.   

\subsection{Astrophysical implications of the model}

The first order Fermi acceleration by the reconnection in a turbulent magnetized fluid may have important astrophysical consequences beyond the Space physics. For instance, interstellar medium observations indicate the existence of the power law from dozens of parsecs to subAU scales (see Crovisier \& Dickey 1983, O'Dell \& Castaneda, Green 1993, Armstrong et al. 1994, Elmegreen \& Scalo 2004, McKee \& Ostriker 2007, Lazarian 2009). Thus it is not inconceivable that magnetic reconnection can play a significant role in the acceleration of cosmic rays on the galactic scale and in other circumstances. We defer a discussion of this interesting possibility for future papers.

Now that the plasma and field data are both available from the Voyagers at the heliosheath, reconnection sites can be directly probed. If it is confirmed that energetic particles observed by Voyagers are accelerated by reconnection, this work could induce further efforts in identifying situations where first order Fermi acceleration arising from the reconnection of the weakly stochastic magnetic field is important. The natural place to look is solar flares. While second order Fermi acceleration is frequently involved to explain energetic particles arising during flares (see La Rosa et al. 2006, Petrosian, Yan \& Lazarian 2007, Yan, Lazarian \& Petrosian 2008), in view of fast reconnection it looks promising that the acceleration is driven by the mechanism we discuss in this paper. The origin of fast ions in solar wind (see Fisk \& Gloeckler 2006), relativistic electrons in the galaxy clusters (see Fusco-Femiano et al. 2004, Raphaeli et al. 2006, Brunetti \& Lazarian 2007) may be also related to the acceleration within reconnection regions.

We also note that magnetic field reversals and reconnection are an intrinsic part of the magnetic field dynamics in accretion disks. For some of these disks, e.g. circumstellar disks, the issues of ionization are important (see Shu et al. 2007). If energetic particles are accelerated in accretion disks, this make magnetic activities there self-sustained. Further research should quantify this and related issues. The theoretical calculations of the reconnection rate in a weakly turbulent partially ionized gas are provided in Lazarian, Vishniac \& Cho (2004).  

\subsection{Summary}

We proposed that the magnetic reconnection could accelerate energetic particles in the heliosheath and especially near the heliopause. This can explain the fact that Voyagers failed to detect the signatures of shock acceleration.  Our predictions include the localization of the source of the energetic particles close to the heliopause. 

\acknowledgments

A.L. thanks Jungyeon Cho, Randy Jokipii and Jim McFadden for elucidating discussions of the properties of the interplanetary current sheets, turbulence and the associated processes of energetic particle acceleration. A.L. acknowledges NSF grants AST 0808118 and ATM 0648699, as well as the support of the NSF Center for Magnetic Self-Organizatin.  The calculation shown are supported by the NSF project TG-AST090078 through TeraGrid resources provided by Texas Advanced Computing Center (TACC:www.tacc.utexas.edu). M.O. would like acknowledge the support of NASA-Voyager Guest Investigator grant NNX07AH20G as well as the support of the National Science Foundation CAREER Grant ATM-0747654. We thank the anonymous referee for very helpful comments and suggestions and Blakesley Burkhart for reading the manuscript. 

\appendix 

\section{Model of MHD Turbulence Adopted}

The nature of Alfvenic cascade is expressed through the critical balance condition in GS95 model of strong turbulence, namely,
\begin{equation} 
\lambda_{\|}^{-1}V_A\sim \lambda_{\bot}^{-1}v_\lambda,
\label{crit}
\end{equation}
 where $v_\lambda$ is the eddy velocity, while the $\lambda_{\|}$ and $\lambda_{\bot}$ are, respectively, eddy scales parallel and perpendicular to the local direction of magnetic field. The critical balance condition states that the parallel size of an eddy is determined by the distance Alfvenic perturbation can propagate during the eddy turnover.The notion of {\it local} is important\footnote{To stress the difference between local and global systems here we do not use the language of $k$-vectors. Wavevectors parallel and perpendicular to magnetic fields can be used, if only the wavevectors are understood in terms of a wavelet transform defined with the local reference system rather than ordinary Fourier transform defined with the mean field system.}, as no universal relations exist if eddies are treated in respect to the global mean magnetic field (LV99, Cho \& Vishniac 2000, Maron \& Goldreich 2001, Lithwick \& Goldreich 2001, Cho, Lazarain \& Vishniac 2002). Combining this with the Kolmogorov cascade notion, i.e. that the energy transfer rate is $v^2_{\lambda}/(\lambda_{\bot}/v_{\lambda})=const$ one gets $\lambda_{\|}\sim \lambda_{\bot}^{2/3}$. If the turbulence injection scale is $L_{inj}$, then $\lambda_{\|}\approx L_{inj}^{1/3}\lambda_{\bot}^{2/3}$, which shows that the eddies get very much anisotropic for small $\lambda_{\bot}$.

The critical balance is the feature of the strong turbulence, which is the case when the turbulent energy is injected at $V_A$. If the energy is injected at velocities lower than $V_A$ the cascade is weak with $\lambda_{\bot}$ of the eddies increasing while $\lambda_{\|}$ staying the same (Ng \& Bhattacharjee 1996,
LV99, Galtier et al. 2002) In other words, as a result of the weak cascade the eddies get thinner, but preserve the same length along the local magnetic field. This decreases $\lambda_{\bot}$ and eventually makes Eq.~(\ref{crit}) satisfied. If the injection velocity is $v_l$ and turbulent injection scale is $l$, the transition to the strong MHD turbulence happens at the scale $l(v_l/V_A)^2$ and the velocity at this scale is $V_{strong}=V_A (v_l/V_A)^2$ (LV99, Lazarian 2006). Thus the weak turbulence has a limited, i.e. $[l, l(v_l/V_A)^2]$ inertial interval and get strong at smaller scales.

While GS95 assumed that the turbulent energy is injected at $V_A$ at the injection scale $l$, LV99 provided general relations for the turbulent scaling at small scales for the case that the injection velocity $v_{l}$ is less of equal to $V_A$, which can be written in terms of $\lambda_{\|}$ and $\lambda{\bot}$:
\begin{equation}
\lambda_{\|}\approx l \left(\frac{\lambda_{\bot}}{l}\right)^{2/3} \left(\frac{V_A}{v_l}\right)^{4/3}
\label{Lambda}
\end{equation}
\begin{equation}
v_{\lambda}\approx v_{l} \left(\frac{l}{\lambda}\right)^{1/3} \left(\frac{v_l}{V_A}\right)^{1/3}
\label{vl}
\end{equation}

The present day debates of whether GS95 approach should be augmented by additional concepts like "dynamical alignment", "polarization", "non-locality" (Boldyrev 2006, 2007, Beresnyak \& Lazarian 2006, 2009,  Gogoberidge 2007) do not change the nature of the reconnection of the weakly turbulent magnetic field as LV99 also considered the modification of reconnection when 
$\lambda_{\|}\sim \lambda_{\bot}^{p} (V_A/v_l)^m$ and showed that for choice of $p$ and $m$ which agrees with the present day simulations the nature of  the reconnection does not  change. 
 

\end{document}